\documentclass[epjCONF,columns]{svjour}
\usepackage{graphics}
\usepackage{amsmath,amssymb}
\usepackage[varg]{txfonts} 
\usepackage[latin1]{inputenc}
\session-title{Hadron Collider Physics Symposium 2011}
\begin{document}
\title{Composite Higgs under LHC Experimental Scrutiny}
\author{J.R.~Espinosa \inst{1}\fnmsep\thanks{\email{espinosa@ifae.es}} \and C.~Grojean\inst{2}\fnmsep\thanks{\email{christophe.grojean@cern.ch}} \and M.~M\"uhlleitner\inst{2}  \fnmsep\thanks{\email{maggie@particle.uni-karlsruhe.de}}}
\institute{ICREA at IFAE, Universitat Aut{\`o}noma de Barcelona, 08193 Bellaterra, Barcelona, Spain \and 
CERN, Physics Department, Theory Unit, CH--1211 Geneva 23, Switzerland \and 
Institute for Theoretical Physics, Karlsruhe Institute of Technology, D-76128 Karlsruhe, Germany}
\abstract{
The LHC has been built to understand the dynamics at the origin of the breaking of the electroweak symmetry. Weakly coupled models with a fundamental Higgs boson have focused most of the attention of the experimental searches. We will discuss here how to reinterpret these searches in the context of strongly coupled models where the Higgs boson  emerges as a composite particle. In particular, we use LHC data to constrain the compositeness scale. We also briefly review the prospects to observe other bosonic and fermionic resonances of the strong sector.
} 
\maketitle

\section{The need for a UV completion of the electroweak Goldstone bosons}
\label{intro}
One important charge of the LHC, if not the main one, is to understand the dynamics responsible for the breaking of the electroweak (EW) symmetry. A massive spin-one particle corresponds to three physical polarizations:  two transverse ones plus an extra longitudinal one which is known to decouple in the massless limit.  Precision EW measurements have established the Brout--Englert--Higgs mechanism and the longitudinal $W^\pm$ and $Z$ certainly correspond to the eaten Nambu--Goldstone bosons associated to the breaking of the global chiral symmetry SU(2)$_L\times$~SU(2)$_R$ to its diagonal subgroup. The gauge field masses are conveniently rewritten as the kinetic terms for these Goldstones ($\sigma^a$, $a=1,2,3$, are the usual Pauli matrices and $v=1/\sqrt{\sqrt{2}G_F}\approx$~246~GeV):
\begin{equation}
\label{eq:pions}
\mathcal{L}_\textrm{\tiny mass}
= \frac{v^2}{4} \textrm{Tr} \left( D_\mu \Sigma ^\dagger D^\mu \Sigma\right)
\ \ 
\textrm{ with} \ \ 
\Sigma=e^{i \sigma^a \pi^a/v},
\end{equation}
which indeed exactly reproduces the standard mass Lagrangian in the unitary gauge ($\Sigma=1$).
However, this is a description that is not self-consistent at very high energy since it leads to scattering amplitudes growing with the energy as it follows from the Goldstone equivalence theorem by expanding the Lagrangian~(\ref{eq:pions}) ($V$ denotes $W^\pm$ or $Z$):
\begin{equation}
\mathcal{A} (V_L^a V_L^b \to V_L^c V_L^d) = \mathcal{A}(s) \delta^{ab}\delta^{cd} + \mathcal{A}(t) \delta^{ac}\delta^{bd} +\mathcal{A}(u) \delta^{ad}\delta^{bc} 
\end{equation}
\begin{equation} 
	\label{eq:LET}
\textrm{ with} \ \ 
\mathcal{A}(s)\approx \frac{s}{v^2}.
\end{equation}
In the absence of any new weakly coupled elementary degrees of freedom canceling this growth, perturbative unitarity will be lost around  $4 \pi v \approx$ 3~TeV (or $2\sqrt{2\pi}v\approx$ 1.2~TeV, depending on the exact criterion used to define strong coupling) and new strong dynamics will kick in and soften the UV behavior of the amplitude, for instance via the exchange of massive bound states similar to the $\rho$ meson of QCD.

\section{Effective Lagrangian for a composite Higgs}

The simplest  example of new dynamics that can restore perturbative unitarity consists of a single (canonically normalized) scalar field, $h$, coupled to the longitudinal $V$'s and to the SM fermions as~\cite{SILH,SILH2,gghh,CLIC}:
\begin{eqnarray}
	\nonumber
&&\hspace{-.2cm}
\mathcal{L}_\textrm{\tiny EWSB}=  \frac{v^2}{4} \textrm{Tr} \left( D_\mu \Sigma ^\dagger D^\mu \Sigma\right) \times \left(1+ 2 a \frac{h}{v} + b \frac{h^2}{v^2} +b_3 \frac{h^3}{v^3} + \ldots \right)\\
	\label{eq:scalar}
&&\hspace{0.5cm}
-\frac{v}{\sqrt{2}} \left( \bar{u}_L^i \bar{d}_L^i \right) \Sigma \left(1+ c \frac{h}{v} +c_2 \frac{h^2}{v^2} \ldots\right) \left( \begin{array}{c} y^u_{ij} u_R^j\\ y^d_{ij} d_R^j \end{array} \right) + h.c.
\end{eqnarray}
The scalar self-interactions can be parametrized as
\begin{equation}
	\label{eq:d3d4}
V(h) = \frac{1}{2} m_h^2 h^2 + d_3 \left( \frac{m_h^2}{2v} \right) h^3 + d_4 \left( \frac{m_h^2}{8v^2} \right) h^4 + \dots
\end{equation}
For $a=1$, the scalar exchange cancels the growing piece~(\ref{eq:LET}) of the $V_LV_L$ amplitude at high energy. 
Furthermore for $b=a^2$, the inelastic $V_L V_L \to hh$ amplitude also remains finite at high energy, while, for $ac=1$ the $V_L V_L \to f \bar{f}'$ amplitude does not grow either. The point $a=b=c=1$ (and also $d_3=d_4=1$, $c_2=b_3=0$, see Table~\ref{tab:abc}) defines the SM Higgs boson and the scalar resonance  then combines with the EW Goldstones to form a doublet transforming {\it linearly} under SU(2)$_L\times$~SU(2)$_R$. When $a\not =1$, the Higgs boson alone fails to fully unitarize the $V_L V_L$ scattering amplitude but the breakdown of pertubative unitarity is pushed to a higher scale of the order of $4\pi v /\sqrt{1-a^2}$. The residual growth of the scattering amplitude $\mathcal{A}(s)\approx (1-a^2) s/v^2$ is ultimately cancelled via the exchange of other degrees of freedom, for instance some vector resonances, which however do not have to be light and could escape any detection at the LHC~\cite{FGKPW}.

Away from the SM point, this set-up (\ref{eq:scalar})--(\ref{eq:d3d4}) introduces minimal deviations in the physics of the Higgs boson: all the Higgs couplings have the same Lorentz structure as in the SM and they are only rescaled by appropriate factors of $a, b$ and $c$ (note that $c$ is flavor-universal and the only source of flavor violation are the usual SM Yukawa couplings, $y^{u,d}$; this minimal flavor violation structure actually emerges naturally in the dynamical models that will be considered below):
\begin{equation}
g_{hVV}=a \, g_{hVV}^{SM}, \ g_{hhVV}=b\,  g_{hhVV}^{SM} \  \ \textrm{and }  \  g_{hf\bar{f}'}=c \, g_{hf\bar{f}'}^{SM}.
\end{equation}
In addition, there are also new couplings, for instance $b_3$ between three Higgses and two gauge bosons or $c_2$ between two Higgses and two fermions, that will contribute to multi-Higgs production~\cite{SILH,SILH2,gghh,CLIC}. 

Since  the NLO QCD corrections do not affect the Higgs couplings,  at the LHC the relevant Higgs  production cross-sections simply rescale as~\cite{SILH3}:
\begin{eqnarray}
\nonumber
&& \resizebox{0.80\columnwidth}{!}{\includegraphics{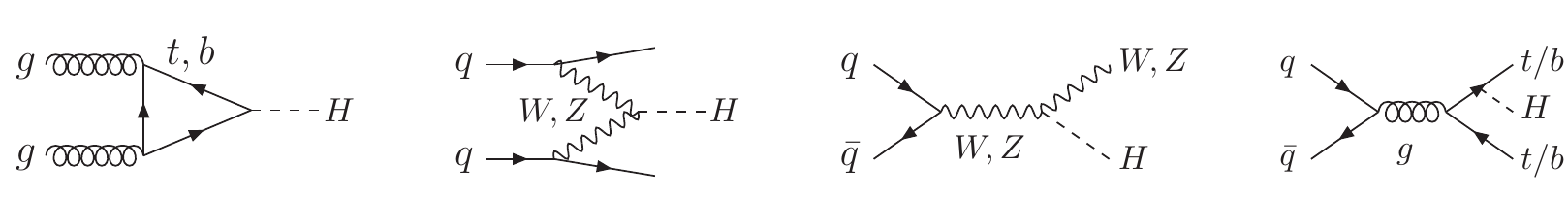}}\\
	\label{eq:xsec}
&\frac{\sigma_{NLO}}{\sigma_{NLO}^{SM}}=
& \hspace{.5cm} c^2 \hspace{1.4cm} a^2  \hspace{1.5cm} a^2  \hspace{1.5cm} c^2
\end{eqnarray}
The loop-induced gluon fusion production could in principle be sensitive to new colored degrees of freedom, e.g. new quarks, running in the loop. But it was shown~\cite{ggfusion} that in explicit Little Higgs models as well as in Composite Higgs models, a delicate cancelation holds and the cross-section is independent of the masses and couplings of these new quarks.

Similarly, the decay widths also have a simple rescaling:
\begin{eqnarray}
\label{eq:Gff} &\Gamma(H\to f\bar{f}) = c^2 \, \Gamma^{SM} (H\to f\bar{f})\, ,\\
\label{eq:GVV} &\Gamma(H\to VV) = a^2 \, \Gamma^{SM}(H\to VV)\, ,\\
\label{eq:Ggg} &\Gamma(H\to gg) = c^2 \, \Gamma^{SM} (H\to gg)\, ,\\
\label{eq:Ggamgam} &\Gamma(H\to \gamma\gamma) = 
\frac{\left( c I_\gamma + a J_\gamma \right)^2}{(I_\gamma +J_\gamma)^2} \Gamma^{SM}(H\to \gamma\gamma)\, ,
\end{eqnarray}
where
\begin{equation}
\begin{array}{c}
I_\gamma=\frac{4}{3} F_{1/2}(4m_t^2/m_h^2), \hspace{.4cm} 
J_\gamma=F_{1}(4m_W^2/m_h^2),\\[.2cm]
F_{1/2}(x)\equiv -2x\left[1+(1-x)f(x)\right],\\[.2cm]
F_{1}(x)\equiv 2+3x\left[1+(2-x)f(x)\right], \\[.2cm]
f(x)\equiv \left\{ \begin{array}{c} 
\arcsin[1/\sqrt{x}]^2\  \mathrm{for}\ x\ge 1\\
-\frac{1}{4} \left[ \log \frac{1+\sqrt{1-x}}{1-\sqrt{1-x}} - i \pi \right]^2\  \mathrm{for}\ x<1
\end{array}
\right.
\end{array}
\end{equation}

\begin{table}
\caption{Values of the couplings of the effective Lagrangian~(\ref{eq:scalar}) in the strongly interacting light Higgs set-up (SILH) and in explicit SO(5)/SO(4) composite Higgs models built in warped 5D space-time (in MHCM$_4$, the SM fermions are embedded into spinoral representations of SO(5) while in MHCM$_5$ they are in fundamental representations). $\xi=(v/f)^2$ measures the amount of compositeness of the Higgs boson.  For the SM with an elementary Higgs, which corresponds to the limit $\xi\to 0$,  the couplings are $a=b=c=d_3=d_4=1$ and $c_2=b_3=0$.}
\label{tab:abc}    
\begin{tabular}{cccc}
\hline\noalign{\smallskip}
Parameters & \hspace{-.5cm} SILH & MCHM4 & MCHM5  \\
\noalign{\smallskip}\hline\noalign{\smallskip}
$a$ &  \hspace{-.5cm}$1-c_H\xi/2$ & $\sqrt{1-\xi}$ & $\sqrt{1-\xi}$\\[.2cm]
$b$ &  \hspace{-.5cm}$1-2 c_H \xi$ & $1-2\xi$ & $1-2\xi$\\[.2cm]
$b_3$ &  \hspace{-.5cm} $-\frac{4}{3}\xi$ & $-\frac{4}{3} \xi \sqrt{1-\xi}$ & $-\frac{4}{3} \xi \sqrt{1-\xi}$ \\[.3cm]
$c$ &  \hspace{-.5cm}$1-(c_H/2+c_y) \xi$ &  $\sqrt{1-\xi}$ & $\frac{1-2\xi}{\sqrt{1-\xi}}$  \\[.3cm]
$c_2$ &  \hspace{-.5cm} $-(c_H+3c_y)\xi/2$ &  $-\xi/2$ & $-2\xi$  \\[.2cm]
$d_3$ &  \hspace{-.5cm}$1+(c_6 - 3 c_H/2) \xi$ & $\sqrt{1-\xi}$ & $\frac{1-2\xi}{\sqrt{1-\xi}}$\\[.3cm]
$d_4$ &  \hspace{-.5cm}$1+(6 c_6 - 25 c_H/3) \xi$ & $1-7 \xi/3$ & $\frac{1-28\xi(1-\xi)/3}{1-\xi}$\\
\noalign{\smallskip}\hline
\end{tabular}
\end{table}

The scalar $h$ could correspond to the usual SM Higgs boson mixed for instance with a gauge singlet  but it  could also be a composite bound state emerging from a strongly interacting sector. When such a composite Higgs boson appears as a fourth Goldstone boson associated to the spontaneous breaking of a global symmetry $G$ of the strong sector to a subgroup $H$, there is a natural mass gap between $f$, the dynamical scale of the strong interactions, i.e. the Goldstone decay constant, and $v$, the electroweak scale that is generated radiatively. These composite Higgs models appear as a natural generalization of the SM with new Goldstones in addition to the $W_L$ and $Z_L$ (see Table~\ref{tab:models}). Without knowing the details of the physics of the strongly interacting theories giving rise to the composite Higgs and other possible resonances, a general effective chiral Lagrangian can capture the low-energy physics of the composite particles~\cite{SILH}. The strong sector is broadly parametrized by two quantities: the typical mass scale, $m_\rho$, of the heavy vector resonances and the dynamical scale, $f$, associated to the global symmetry pattern $G/H$. The effective chiral Lagrangian includes only four operators that are genuinely sensitive to the strong interactions and affect qualitatively the physics of the strongly interacting light Higgs (SILH) boson:
\begin{equation}
 \label{eq:silh}
\begin{array}{c}
\hspace{-.5cm}
\mathcal{L}_{\rm SILH} = \frac{c_H}{2f^2} \left( \partial_\mu \left( H^\dagger H \right) \right)^2
+ \frac{c_T}{2f^2}  \left(   H^\dagger{\overleftrightarrow D}_\mu H\right)^2 \\
\hspace{1cm}
- \frac{c_6\lambda}{f^2}\left( H^\dagger H \right)^3 
+ \left( \frac{c_y y^{ij}_f}{f^2}H^\dagger H  {\bar f}^i_L H f^j_R +{\rm h.c.}\right) 
\end{array}
\end{equation}
Whenever this chiral Lagrangian emerges from a strong sector that is invariant under a custodial symmetry, the coefficient $c_T$ vanishes.
The values of the couplings $a, b, \ldots$ obtained from this SILH Lagrangian are given in Table~\ref{tab:abc}. The SILH Lagrangian can be extended in several ways (see Refs.~\cite{FGKPW,resonances}) to include some heavy vector resonances of the strong sector in addition to the Goldstone bosons.

\begin{table}
\caption{Global symmetry breaking patterns and the corresponding Goldstone boson contents of the SM, the minimal composite Higgs model, the next to minimal composite Higgs model, the minimal composite two Higgs doublet model. Note that the SU(3) model does not have a custodial invariance. $a$ denotes a CP-odd scalar while $h$ and $H$ are CP-even scalars}
\label{tab:models}    
\begin{tabular}{ccc}
\hline\noalign{\smallskip}
Model & Symmetry Pattern & Goldstones   \\
\noalign{\smallskip}\hline\noalign{\smallskip}
SM & SO(4)/SO(3) & $W_L, Z_L$\\
--- & SU(3)/SU(2)$\times$U(1) & $W_L, Z_L, h$\\
MCHM & SO(5)/SO(4)$\times$U(1) & $W_L, Z_L, h$\\
NMCHM & SO(6)/SO(5)$\times$U(1) & $W_L, Z_L, h, a$\\
MCTHM & SO(6)/SO(4)$\times$SO(2) $\times$U(1) & $W_L, Z_L, h, H, H^\pm, a$\\
\noalign{\smallskip}\hline
\end{tabular}
\end{table}

\section{Explicit 5D composite Higgs models}

\begin{figure}[t]
\begin{center}
\resizebox{0.95\columnwidth}{!}{\includegraphics{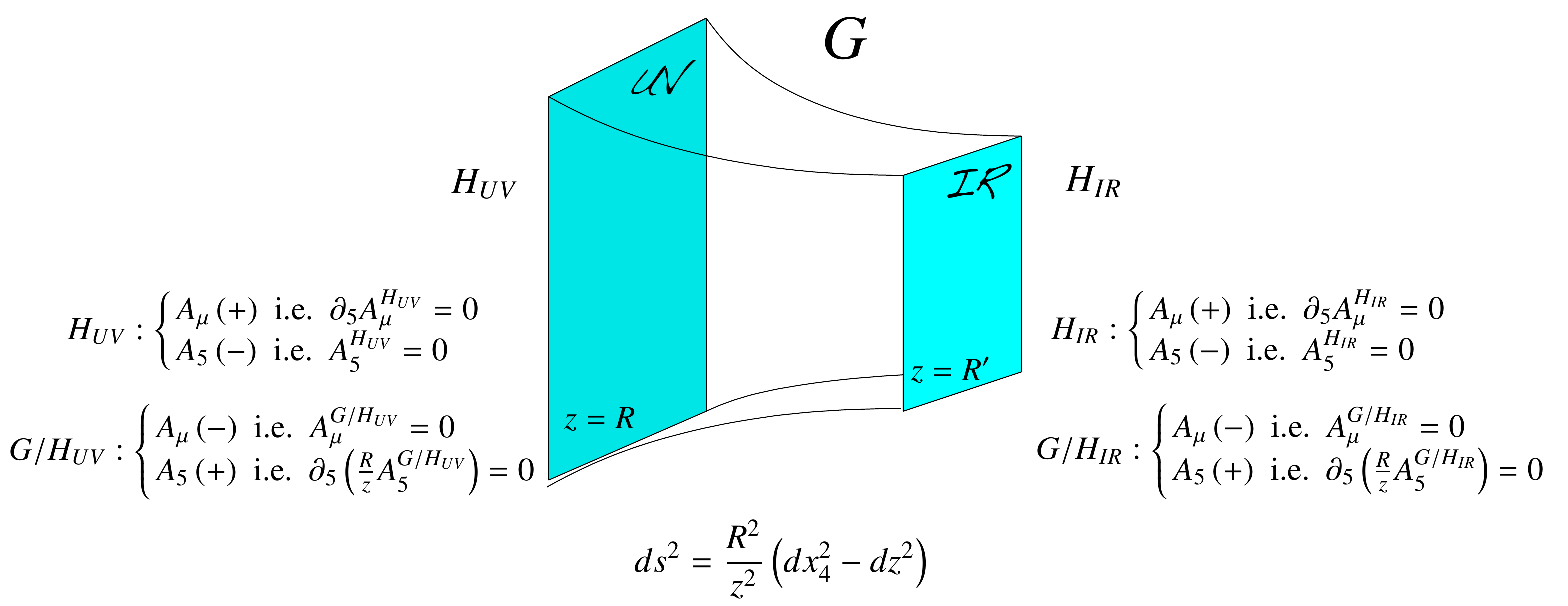}}\\
\resizebox{0.40\columnwidth}{!}{\includegraphics{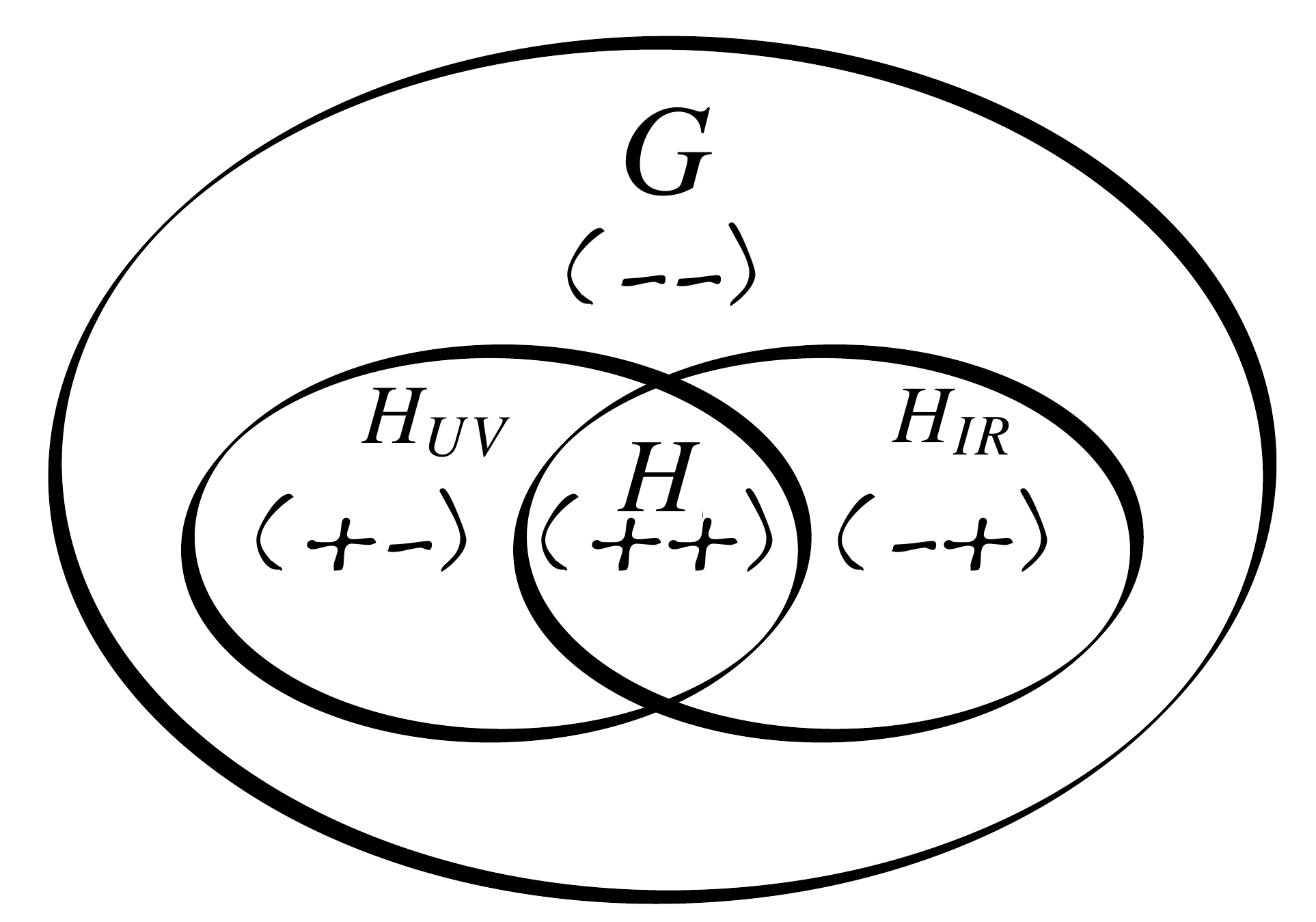}}\\
\caption{Composite models built in five dimensional Anti-de-Sitter space-time and their symmetry breaking pattern interpretation. In 5D, the gauge symmetry in the bulk, $G$, is broken by suitable boundary conditions to $H_{UV}$ on the UV brane and to $H_{IR}$ on the IR brane. The low energy theory mimics in 4D a strongly interacting sector invariant under a global symmetry $G$ spontaneously broken to $H_{IR}$ at the IR scale with a subgroup $H_{UV}$ which is weakly gauged. The number of Goldstone bosons is equal to $\textrm{dim}(G/H_{IR})$,  $\textrm{dim}(H_{UV}/H)$ being eaten to give a mass to some gauge bosons ($H=H_{UV} \cap H_{IR}$). The remaining $\textrm{dim}(G/H_{IR})-\textrm{dim}(H_{UV}/H)$ massless Goldstones are described on the 5D side by the massless $A_5^{H}$ modes.}
 \label{fig:AdSCFT}
\end{center}
\end{figure}

The SILH Lagrangian should be seen as an expansion in $\xi = (v/f)^2$. It can therefore be used in the vicinity of the SM limit ($\xi \to 0$), whereas the technicolor limit ($\xi \to1$), when the scale of the strong interaction becomes degenerate with the weak scale, requires a resummation of the full series in 
$\xi$. Explicit models provide concrete examples of such a resummation. Here we refer to the Holographic Higgs models of Refs.~\cite{MCHM5D}, which are based on a five-dimensional gauge theory in Anti-de-Sitter (AdS) space-time. Via the AdS/CFT correspondence, these 5D models describe 4D strongly coupled models, the components of the gauge fields along the fifth dimension being interpreted as the Goldstone bosons of the strong sector (see Fig.~\ref{fig:AdSCFT}).
The minimal models are based on an SO(5)$\times$U(1) symmetry in the bulk, broken to the SM SU(2)$_L \times$~U(1)$_Y$ gauge group on the UV brane and to  SO(4)$\times$~U(1) on the IR brane. These models contain exactly four massless $A_5$ degrees of freedom which transform as a doublet of SU(2)$_L$ and whose wave-functions are peaked on the IR. Radiative corrections generate a {\it finite} potential for these $A_5$ degrees of freedom, the finiteness results in 5D from the fact that gauge invariance forbids any local term and the potential emerges from non-local effects along the fifth dimension. The 4D interpretation is simply that the Higgs doublet is a composite field and the radiative corrections are screened by the compositeness scale. The exact form of the potential depends on the way the SM fermions are embedded in representations of SO(5): when they belong to spinorial representations (MCHM4), the potential is of the form
\begin{equation}
V(h)=\alpha \cos (h/f) - \beta \sin^2 (h/f),
\end{equation}
which after expanding around the EW vacuum leads to the particular values of $d_3$ and $d_4$ reported in Table~\ref{tab:abc}.
When the SM fermions are part of fundamental representations of SO(5) (MCHM5), the Higgs potential takes the form:
\begin{equation}
V(h)=\alpha \sin^2 (h/f) - \beta \sin^2 (2 h/f),
\end{equation}
leading to different values of $d_3$ and $d_4$.

Both in MCHM4 and in MCHM5, the quadratic terms in the $W$ and $Z$ bosons read:
\begin{equation}
	\label{eq:MCHMgauge}
m^2_W(h) 
\left( W_\mu 
W^\mu+\frac{1}{2\cos^2\theta_W}
Z_\mu Z^\mu\right)
\end{equation}
with $m_W(h)=\frac{gf}{2}\sin\frac{h}{f}$.
Again expanding around the EW vacuum, we obtain the expression of the weak scale
\begin{equation}
v= f \sin (\langle h\rangle/f),
\end{equation}
and the values of the coefficients $a$ and $b$ reported in Table~\ref{tab:abc}.

In MCHM4, the interactions of the Higgs to the fermions take the form
\begin{equation}
{\cal L}_{\rm Yuk}=-m_f(h){\bar f}f \hspace{.2cm} \textrm{ with } 
\hspace{.2cm}  
m_f(h)= M \sin(h/f).
\label{eq:MHCM4}
\end{equation}
We then obtain 
\begin{equation}
\textrm{MCHM4:} \hspace{.2cm} c=\sqrt{1-\xi} \ \ \textrm{and} \ \ c_2=-\xi/2.
\end{equation}
In MCHM5, the interactions of the Higgs to the fermions is changed to
\begin{equation}
{\cal L}_{\rm Yuk}=-m_f(h){\bar f}f \hspace{.2cm} \textrm{ with } 
\hspace{.2cm}  
m_f(h)= M \sin(2h/f),
\label{eq:MHCM5}
\end{equation}
and therefore 
\begin{equation}
\textrm{MCHM5:} \hspace{.2cm} c=(1-2\xi)/\sqrt{1-\xi} \ \ \textrm{and} \ \ c_2=-2\xi.
\end{equation}
Note that in MHCM5, the Higgs boson becomes fermiophobic for $\xi=0.5$.

\section{Direct and indirect constraints on composite Higgs models}

\subsection{EW precision data constraints}
\begin{figure}[t]
\begin{center}
\resizebox{0.95\columnwidth}{!}{\includegraphics{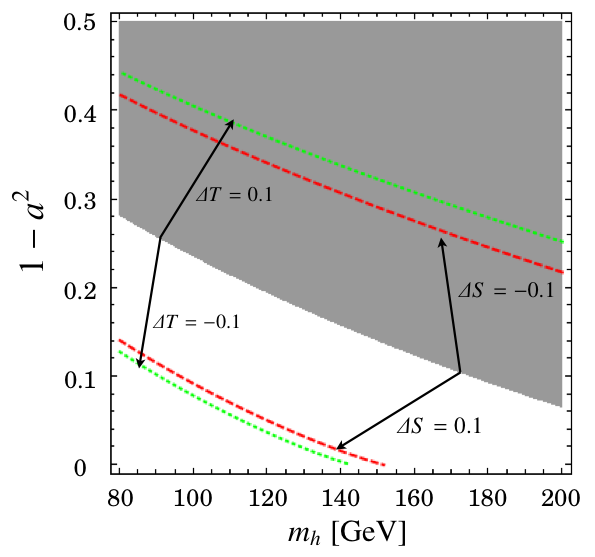}}
\caption{Limits from EW precision data. The grey area is excluded at 99\% CL when no additional contributions to $S$ and $T$ beyond the ones of Eq.~(\ref{eq:ST}) are introduced. The dashed-red (dotted green) lines indicate how this exclusion area is modified in the presence of additional  contributions to $S$ ($T$). The logarithmic divergences are cut-off at $\Lambda=4 \pi v/\sqrt{1-a^2}$. The usual parameters $\epsilon_2$ and $\epsilon_b$ used in EW fits are kept fixed to their SM values.
}
 \label{fig:EWdata}
\end{center}
\end{figure}

Within the SM, the Higgs boson ensures the proper decoupling of the longitudinal polarization of the massive gauge bosons at high energy.
But it is also there to screen the radiative corrections to the propagators of the transverse polarizations which otherwise will be logarithmically divergent. This screening is the result of a cancelation between a gauge-boson loop and a Higgs loop. Clearly this cancelation does not hold any longer when the Higgs couplings are modified, more precisely, when $a$, the linear coupling to gauge fields, is different than its SM value. This translates into a logarithmic contribution to the oblique parameters $S$ and~$T$~\cite{HiggsIR}:
\begin{equation}
	\label{eq:ST}
\begin{array}{c}
\Delta S  \approx  \frac{1}{6\pi}  \left(\log(m_h/m_Z) -(1-a^2)
\log (m_h/\Lambda) \right)\\[.2cm]
\Delta T  \approx -\frac{3}{8 \pi c^2_W} \left(\log(m_h/m_Z) - (1-a^2)
\log (m_h/\Lambda) \right)
\end{array}
\end{equation}
The usual fit of EW data can then be used to constrain the $(m_h,a)$ plane, see for instance Fig.~\ref{fig:EWdata}.
For a given Higgs mass, the allowed range  of values of the parameter $a$ depends on possible additional contributions to $S$ and $T$ from UV physics, like a positive contribution to $S$ from heavy vector resonances and positive or negative contributions from new quarks.

\subsection{Direct searches}

\begin{figure}[t]
\begin{center}
\resizebox{0.95\columnwidth}{!}{\includegraphics{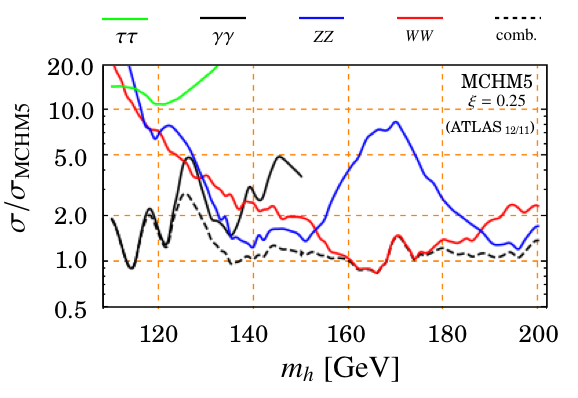}}
\resizebox{0.95\columnwidth}{!}{\includegraphics{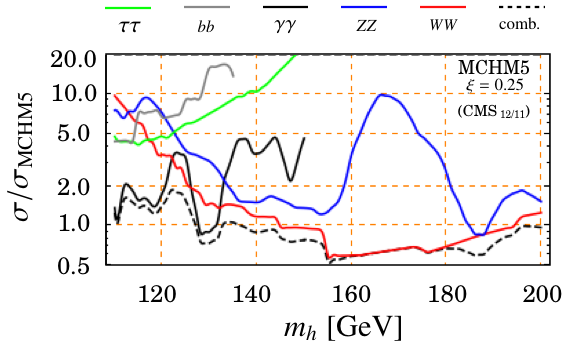}}
\caption{MCHM5: Rescaled limits from the search of a Higgs boson in various channels set by the ATLAS (top) and CMS (bottom) data presented at CERN in December 2011~\cite{Dec11}. The various channels are added in quadrature.}
 \label{fig:MHCM5ratiosigma}
\end{center}
\end{figure}

\begin{figure}[t]
\begin{center}
\resizebox{0.95\columnwidth}{!}{\includegraphics{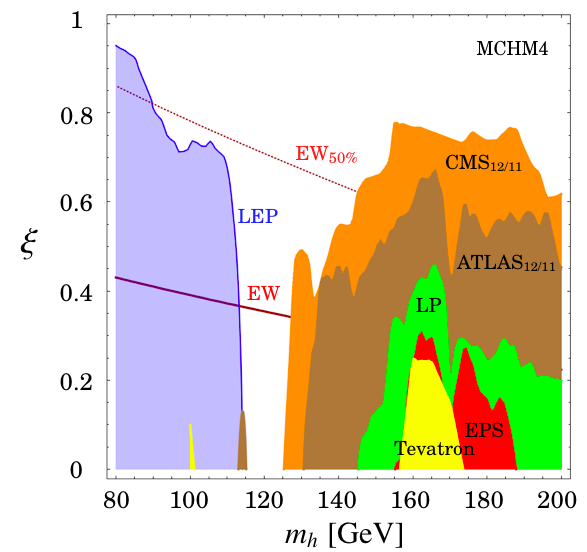}}
\caption{Limits from Higgs searches at LEP, Tevatron and LHC in the plane ($m_h, \xi$) for MCHM4. For the LHC constraints, we have used the data reported at the EPS-HEP 2011 conference~\cite{EPS}, the Lepton-Photon 2011 symposium~\cite{LP} and the ones announced in December 2011 at CERN~\cite{Dec11}. The individual channels are appropriately rescaled according to Eqs.~(\ref{eq:xsec})--(\ref{eq:Ggamgam}) and they are then simply combined in quadrature. The red continuous line delineates the region favoured at 99\% CL by EW precision data (with a cutoff scale of 2.5 TeV and after marginalizing over $\epsilon_2$ and $\epsilon_b$), the region below the red dashed line survives for an additional 50\% cancellation of the oblique parameters.}
 \label{fig:MHCM4searches}
\end{center}
\end{figure}

\begin{figure}[t]
\begin{center}
\resizebox{0.95\columnwidth}{!}{\includegraphics{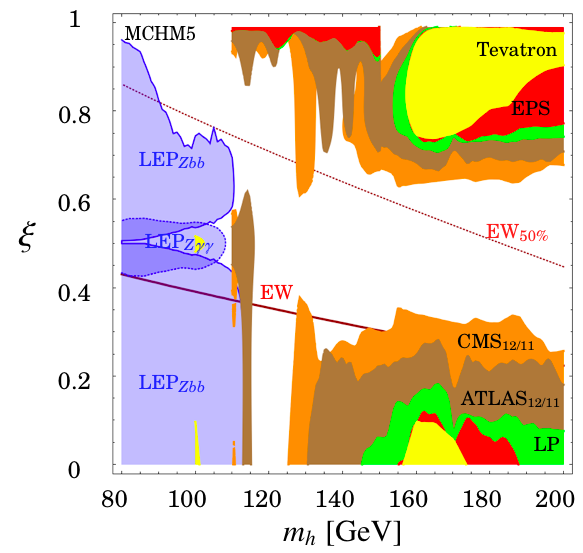}}
\caption{Same as Fig.~\ref{fig:MHCM4searches} but for MCHM5.}
 \label{fig:MHCM5searches}
\end{center}
\end{figure}

In explicit models, the Higgs production cross-sections and decay branching-ratios are simple functions of the parameter $\xi$ that measures the amount of compositeness of the Higgs boson. Therefore, the searches for the SM Higgs boson can be trivially rescaled with the parameter $\xi$ once we notice that only the signal rates and not the background ones are changed by the parameter $\xi$, see Fig.~\ref{fig:MHCM5ratiosigma}. Stringent exclusion bounds in the 2D parameter space $(m_h,\xi)$ are obtained when the LHC various channels~\cite{Dec11} are combined. Figures~\ref{fig:MHCM4searches} and~\ref{fig:MHCM5searches} present the results when the various channels are simply added in quadrature. A more proper statistical combination of the individual channels leads to similar results~\cite{Contino}.

The Higgs searches at the LHC have a chance to establish a deviation away from the SM. 
Nonetheless, measuring $a \not =1 $ would not necessarily establish that EW symmetry breaking is triggered by a new strong force since possible weakly coupled deformations of the SM could also lead to Higgs anomalous couplings.
The true nature of the strong interactions would be revealed in some particular processes like $V_L V_L \to V_L V_L$~\cite{SILH} or $V_L V_L \to hh$~\cite{SILH2}. Other processes like $V_L V_L \to hhh$~\cite{CLIC} could even teach us about some discrete symmetries of the strong interactions.

\subsection{Direct searches of other resonances}

\begin{figure}[t]
\begin{center}
\resizebox{0.45\columnwidth}{!}{\includegraphics{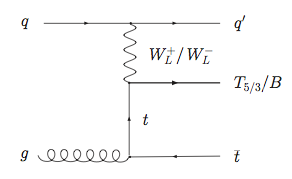}}
\hspace{.2cm}
\resizebox{0.45\columnwidth}{!}{\includegraphics{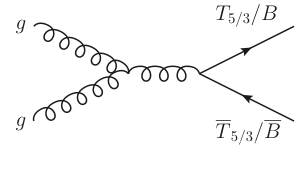}}\\
\resizebox{0.80\columnwidth}{!}{\includegraphics{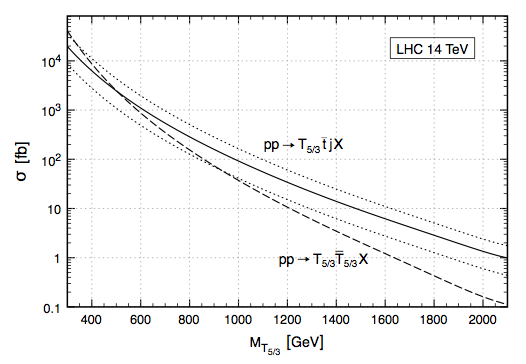}}\\
\caption{Single (top left) and pair (top right) production of resonances in the top sector. The production cross section (bottom) makes a discovery plausible even with a limited integrated luminosity. From Ref.~\cite{fermions}}
 \label{fig:toppartners}
\end{center}
\end{figure}

Of course the direct observation of other states belonging to the strong sector, like vector resonances~\cite{FGKPW,resonances} or some fermionic composite particles~\cite{fermions}, would be tantalizing  signatures auguring a new era in the history of high-energy physics. Quite generically, the vector resonances have only tiny couplings to the light SM quarks via their mixing to the SM EW gauge bosons and therefore their main decay channels are into a pair of longitudinal $W^\pm_L, Z_L$ or a longitudinal $W^\pm_L, Z_L$ together with a Higgs boson. The current most stringent limit of such vector resonances is coming from the search for $WZ$-resonances at the LHC via DY production after mixing with a $W^*$ boson~\cite{CMS-EXO-11-041-pas}\\
\centerline{\resizebox{0.95\columnwidth}{!}{\includegraphics{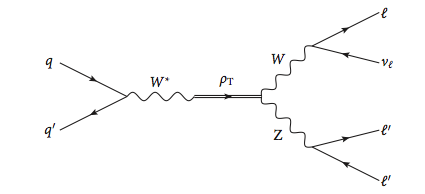}}}
and constrains their masses to be above around 950~GeV for $\xi=0.1$~\cite{FGKPW,CLIC}. The 14~TeV run could push this bound up to 1.5~TeV~\cite{CLIC}. Even if more model-dependent, the production of fermionic resonances in the top sector is quite promising as the large top mass generically implies the existence of light fermionic resonances, possibly the lightest non-Goldstone particles of the strong sector~\cite{fermions,Pomarol:2008bh}.
These fermionic resonances have an exotic 5/3 electric charge and can be produced in pairs or alone, see Fig.~\ref{fig:toppartners}, with a cross section large enough to make a discovery plausible even with a limited integrated luminosity~\cite{fermions}. The existence of these light fermionic states could also be inferred indirectly through the modification they induce in Higgs physics. If their effects on $gg\to h$ actually cancel~\cite{ggfusion}, they give sizable contributions to $gg \to hh$~\cite{gghhNEW}. 
 
\section{Conclusion}
The LHC has been built to understand the dynamics at the origin of the breaking of the electroweak symmetry. The SM Higgs boson, namely a fundamental scalar field with appropriate couplings to the SM particles, is a very compelling UV completion to the Higgs mechanism which ensures the proper decoupling at high energy of the extra polarization associated to the masses of the EW gauge bosons, and at the same time also screens the radiative corrections to the propagators of these $W$ and $Z$. The main virtue of this scenario is that it can remain perturbative, hence calculable, up to very high energy, possibly the GUT or the Planck scale. Nonetheless, strong dynamics could also be responsible for the EW breaking. And surprisingly enough, these strong models can be very similar to the SM if a scalar field with the same quantum number as the Higgs boson emerges as composite bound state from the strong sector.  Viewed from the low-energy perspective, this new setup will appear as a deformation of the SM itself.

\section*{Acknowledgments}
CG would like to thank the organizers of the HCP symposium for their invitation to present our results. This work has been partly supported by the European Commission under the contract ERC advanced grant 226371 `MassTeV', the contract PITN-GA-2009-237920 `UNILHC', and the contract MRTN-CT-2006-035863 `ForcesUniverse', as well as by the Spanish Consolider Ingenio 2010 Programme CPAN (CSD2007-00042) and the Spanish Ministry MICNN under contract FPA2010-17747 and FPA2008-01430. MM is supported by the DFG SFB/TR9 `Computational Particle Physics'.


\begin{thebibliography}{}

\bibitem{SILH}
G.~F.~Giudice, C.~Grojean, A.~Pomarol and R.~Rattazzi,
  JHEP {\bf 0706} (2007) 045
  [hep-ph/0703164].

\bibitem{SILH2}
R.~Contino, C.~Grojean, M.~Moretti, F.~Piccinini and R.~Rattazzi,
  JHEP {\bf 1005} (2010) 089
  [arXiv:1002.1011 [hep-ph]].

\bibitem{gghh}
R.~Grober and M.~Muhlleitner,
  JHEP {\bf 1106} (2011) 020
  [arXiv:1012.1562 [hep-ph]].

\bibitem{CLIC}
R.~Contino, C.~Grojean, D.~Papadopulo, R.~Rattazzi and A.~Thamm,
{\it in preparation}.

\bibitem{FGKPW}
  A.~Falkowski, C.~Grojean, A.~Kaminska, S.~Pokorski and A.~Weiler,
  JHEP {\bf 1111} (2011) 028
  [arXiv:1108.1183 [hep-ph]].

\bibitem{SILH3}
J.~R.~Espinosa, C.~Grojean and M.~Muhlleitner,
  JHEP {\bf 1005} (2010) 065
  [arXiv:1003.3251 [hep-ph]].

\bibitem{ggfusion}
A.~Falkowski,
  Phys.\ Rev.\ D {\bf 77} (2008) 055018
  [arXiv:0711.0828 [hep-ph]].
I.~Low and A.~Vichi,
  Phys.\ Rev.\ D {\bf 84} (2011) 045019
  [arXiv:1010.2753 [hep-ph]].
A.~Azatov and J.~Galloway,
  arXiv:1110.5646 [hep-ph].


\bibitem{resonances}
G.~Panico and A.~Wulzer,
  JHEP {\bf 1109} (2011) 135
  [arXiv:1106.2719 [hep-ph]].
  S.~De Curtis, M.~Redi and A.~Tesi,
  arXiv:1110.1613 [hep-ph].
  R.~Contino, D.~Marzocca, D.~Pappadopulo and R.~Rattazzi,
  JHEP {\bf 1110} (2011) 081
  [arXiv:1109.1570 [hep-ph]].
  F.~Bernardini, F.~Coradeschi and D.~Dominici,
  arXiv:1111.4904 [hep-ph].

\bibitem{MCHM5D}
R.~Contino, Y.~Nomura and A.~Pomarol,
  Nucl.\ Phys.\ B {\bf 671} (2003) 148
  [hep-ph/0306259].
  K.~Agashe, R.~Contino and A.~Pomarol,
  Nucl.\ Phys.\ B {\bf 719} (2005) 165
  [hep-ph/0412089].
  R.~Contino, L.~Da Rold and A.~Pomarol,
  Phys.\ Rev.\ D {\bf 75} (2007) 055014
  [hep-ph/0612048].

\bibitem{HiggsIR}
R.~Barbieri, B.~Bellazzini, V.~S.~Rychkov and A.~Varagnolo,
  Phys.\ Rev.\ D {\bf 76} (2007) 115008
  [arXiv:0706.0432 [hep-ph]].

\bibitem{Dec11}
ATLAS Collaboration, ATLAS--CONF--2011--163.
CMS Collaboration, CMS--PAS--HIG--11--032.



\bibitem{Contino}
A.~Azatov, R.~Contino, J.~Galloway,
{\it in preparation}

\bibitem{fermions}
R.~Contino and G.~Servant,
  JHEP {\bf 0806} (2008) 026
  [arXiv:0801.1679 [hep-ph]].
 J.~Mrazek and A.~Wulzer,
  Phys.\ Rev.\ D {\bf 81} (2010) 075006
  [arXiv:0909.3977 [hep-ph]].
  G.~Panico and A.~Wulzer,
  JHEP {\bf 1109} (2011) 135
  [arXiv:1106.2719 [hep-ph]].

\bibitem{EPS}
ATLAS Collaboration, ATLAS-CONF--2011--112.
CMS Collaboration, CMS--PAS--HIG--11--011.

\bibitem{LP}
ATLAS Collaboration, ATLAS--CONF--2011--135.
CMS Collaboration, CMS--PAS--HIG--11--022.


\bibitem{CMS-EXO-11-041-pas}
CMS Collaboration, CMS--PAS--EXO--11--041.
  
\bibitem{Pomarol:2008bh}
  A.~Pomarol and J.~Serra,
  Phys.\ Rev.\ D {\bf 78} (2008) 074026
  [arXiv:0806.3247 [hep-ph]].

\bibitem{gghhNEW}
J.R.~Espinosa, M.~Gillioz, C.~Grojean, R.~Grober, M.~Muhlleitner and E.~Salvioni,
{\it in preparation}.

\end{thebibliography}
\end{document}